\documentclass[preprint,showpacs,amsmath,amssymb,eqsecnum]{revtex4}
\usepackage{dcolumn}
\usepackage{bm}
\usepackage{graphicx}

\begin{document}

%\draft
\title{{Majorana Neutrinos in Muon Decay}\footnote{Talk given at 
NNR05 workshop on Neutrino Nuclear Responses in Double Beta Decays
and Low-energy Astro-neutrinos, CAST and SPring-8, Japan, 2-4 December 2005.}
} 

\author{Hiroyuki NISHIURA}
\email{nishiura@is.oit.ac.jp}
\affiliation{%
Faculty of Information Science and Technology, 
Osaka Institute of Technology, %\\
Hirakata, Osaka 573-0196, Japan \\ \\ \\}

%\date{January 27, 2006}

\begin{abstract}
The normal muon decay $\mu^{+} \to e^{+}\nu_{e}\overline{\nu_{\mu}}$ 
is studied 
as a tool to discriminate between the Dirac and Majorana type neutrinos. 
For the purpose to do this, we propose a new parameterization in 
place of the Michel parameters for the energy spectrum of $e^{+}$. 
The $\chi^2$-fitting is used by noticing different energy spectra 
between the Dirac and Majorana neutrino cases. 
We assume the interaction Hamiltonian 
which consists of the $V-A$ and $V+A$ charged currents. 
\end{abstract}

%%%%%%%%%%%%%%%%%%%%%%%%%%%%%%%%%%%%%%%%%%%%%%%%%%%%%%%%%%%%%%%%%%%%%%%%%%%%

\maketitle

\section{Introduction}
Neutrinoless double beta decay~\cite{Ejiri} is a best subject for 
investigating whether 
the neutrino is of the Dirac or Majorana type. 
However, it needs to improve the present sensitivity of the detector 
for observing this decay mode, because of the tiny neutrino masses 
and/or the small contributions from the right-handed weak current.
Under this circumstances, it seems to be meaningful as a complementary study 
to survey whether muon decay can be used as a tool to determine the type of the 
neutrino. Recently, we have proposed a new parameterization of 
the $e^{\pm}$ energy spectrum of the muon decay that is suitable 
for discriminating between the types of neutrino~\cite{Doi2005}. 
It is shown that there is a possibility to take advantage of the 
different energy dependences for their energy spectra. 
We propose a method in which the $\chi^{2}$ value determined 
experimentally by assuming the Dirac-type neutrino is compared with 
the one determined for the Majorana-type neutrino. 
This may provide a test to determine the type of neutrino, although it 
is indirect.

\section{General framework}
We assume the following effective weak interaction Hamiltonian for the 
$\mu^{\pm}$ decay~\cite{Doi1} 
\begin{equation}
{\cal{H}}_W(x)=\frac{G_F}{\sqrt{2}}\{ 
    j_{eL\, \alpha}^\dagger j_{\mu L}^\alpha
    + \lambda j_{eR\,  \alpha}^\dagger j_{\mu R}^\alpha
    + \eta j_{eR\,  \alpha}^\dagger j_{\mu L}^\alpha
    +\kappa j_{eL\,  \alpha}^\dagger j_{\mu R}^\alpha
    \} +\mbox{H.c.}\ ,
    \label{Hamiltonian}
\end{equation}
where $G_{F}$ is the Fermi coupling constant. 
This weak interaction is naturally 
expected from the gauge models that contain both $V-A$ and $V+A$ currents 
with the left- and right-handed weak gauge bosons.
The appearance of the coupling constant $\lambda$ is mainly due to the 
right-handed weak gauge bosons $W_{R}$, while terms with $\eta$ and 
$\kappa$ come from the possible mixing between the left-handed  and 
right-handed weak gauge bosons, $W_{L}$ and $W_{R}$. 
In the $SU(2)_L \times SU(2)_R \times U(1)$ gauge model, 
the coupling constants $\kappa$ and $\eta$ become identical. 
However, they are treated as independent constants in this note in 
order to 
allow comparison with the more general case without a restriction 
from the gauge theory (see, e.g., \cite{Fetscher}). 
\par
The left-handed and right-handed charged weak leptonic currents, 
$j_{\ell L}$ and $j_{\ell R}$ with flavor $\ell=e$ and $\mu$, are 
expressed in terms of the mass eigenstates of charged 
leptons $E_{\ell}$ and neutrinos $N_j$ with mass $m_j$ as
\begin{equation}
 j_{\ell L\, \alpha}
  = \sum_{j=1}^{2n}\overline{{E_{\ell}}}\gamma_\alpha(1-\gamma_5)
    U_{\ell j}N_j,
\quad
 j_{\ell R\, \alpha}
  = \sum_{j=1}^{2n}\overline{{E_{\ell}}}\gamma_\alpha(1+\gamma_5)
    V_{\ell j}N_j,
    \label{eq:2no03}
\end{equation} 
for the case of the $n$ generations. 
Here $U_{\ell j}$ and $V_{\ell j}$ stand for the left-handed 
and right-handed lepton mixing matrices, respectively.

%%%%%%%%%%%%%%%%%%%%%%%%%%%%%%%%%%%%%%%%%%%%%%%%%%%%%%%%%%%%%%%%%%%%%%%%%%%%
\section{Differential decay rate for normal muon decay}
The $\mu^{\pm}$ decay takes place as
\begin{equation}
\mu^{\pm} \rightarrow e^{\pm} \ +\ N_j \ +
    \ \overline{N_k},
    \label{eq:2no04} 
\end{equation}
where $\overline{N_k}$ represents an antineutrino for the 
Dirac neutrino case, but it should be understood as 
$N_k$ for the Majorana neutrino case.
\par
If the radiative corrections are not 
included, the differential 
decay rate for 
emitted positron in the 
rest frame of polarized $\mu^{+}$ is expressed 
as~\cite{Doi2,Fetscher}
\begin{equation}
\frac{d^2 \Gamma(\mu^{+} \rightarrow e^{+}\nu \overline{\nu})}
     {d x \, d \cos \theta}
     =\left( \frac{m_{\mu} \, G_{F}^2 \, W^4}
     {6 \cdot 4 \, (\pi)^3} \right)
     \, A\,\sqrt{x^2 - x_{0}^2} \, D(x, \, \theta) \, ,
       \label{eq:2no05}
\end{equation}
where 
\begin{equation}
     x  =  \frac{E}{W}, \, 
     x_{0} =  \frac{m_{e}}{W} = 9.65 \times 10^{-3} , \,\,
     W  = \frac{\, m_{\mu}^{2} + m_{e}^{2} \,}
     {2 \,  m_{\mu}}  = 52.8 \,\, \mbox{MeV}.
     \label{eq:2no06}
\end{equation}
Here $m_\mu$ and $m_e$ are the muon and electron masses, respectively, 
and $E$ is the energy of $e^{+}$. 
The angle $\theta$ represents the direction of the emitted $e^{+}$ 
with respect to the muon polarization vector 
$\vec{P_{\mu}}$ at the instant of the $\mu^{+}$ decay. 
The allowed range of $x$ is limited kinematically as 
\begin{equation}
    x_{0} \leqq x \leqq x_{\mathrm{max}}=(1 - r_{jk}^{2}) \simeq 1
    \quad \mbox{with} \quad r_{jk}^{2} = 
    \frac{(m_{j} + m_{k})^{2}}{2 m_{\mu} W}.
\end{equation}
Here $m_j$ and $m_k$ are masses of neutrinos emitted in the $\mu^{+}$ decay. 
\par
The constant $A$ in Eq.~(\ref{eq:2no05}) is introduced to simplify the 
expression for the energy spectrum 
by taking the arbitrariness of its normalization. 
It is referred to as a normalization factor.  
There are various possibilities for the choice 
of $A$, when experimental data are 
analyzed, although these choices differ only by rearrangements 
of the terms in the theoretical expression.  See Appendix A.

%%%%%%%%%%%%%%%%%%%%%%%%%%%%%%%%%%%%%%%%%%%%%%%%%%%%%%%%%%%%%%%%%%%%%%%%%%%%
\section{Energy spectrum of positron}
We ignore some small terms proportional to both $m_e$ and neutrino 
masses ($m_j$ and $m_k$) in this note, for simplicity. 
Then, the $e^{+}$ energy spectrum part, $xD(x, \, \theta)$, is expressed 
as\\
\begin{equation}
xD(x, \, \theta) = x[ \, N(x) 
      + P_{\mu} \, \cos \theta \, P(x) \, ],
      \label{eq:2no07}
\end{equation}
where $P_{\mu}=|\vec{P_{\mu}}|$ is the rate of muon polarization, 
and the isotropic part $N(x)$ and anisotropic part $P(x)$ are
\begin{eqnarray}
N(x)&=&  \frac{1}{A} 
  \left[ a_{+} (3 x - 2 x^{2} )
     + 12 ( \, k_{+ \, c} + \varepsilon_{m} \, k_{+ \, m} \, )
     \, x \, (1 - x) \right],      \label{eq:3no01} \\
P(x)&=& \frac{1}{A} 
     \left[ a_{-} (-x + 2 \, x^2 )
     + 12 \, ( \, k_{- \, c} + \varepsilon_{m} k_{- \, m} \, )
     \, x \, (1 - x) \right] .
     \label{eq:3no02}
\end{eqnarray}
Here, the decay formulae for the Dirac and Majorana
neutrinos are obtained by setting $\varepsilon_{m} = 0$ 
and $\varepsilon_{m} = 1$, respectively.  
The first terms with setting $A=a_{+}=a_{-}=1$ in these $N(x)$ and 
$P(x)$ correspond to the theoretical predictions from the standard model.
The well-known Michel parameterization~\cite{Fetscher} is obtained if we 
choose the normalization factor 
$A = A_{1 \, 0} = a_{+} + 2 \, k_{+ \, c}$. 
%%%%%%%%%%%%%%%%%%%%%%%%%%%%%%%%%%%%%%%%%%%%%%%%%%%%%%%%%%%%%%%%%%%%%%%%%%%%
\section{Coefficients}
\par
In the Dirac neutrino case, the coefficients in Eqs.(\ref{eq:3no01}) 
and (\ref{eq:3no02}) are expressed as follows:
\begin{eqnarray}
a_{\pm} = \left( 1 \pm \lambda^2 \right)
     \hspace{3mm} \mbox{and} \hspace{3mm}
k_{\pm \, c} =% \frac{1}{2} 
     \left( \kappa^2 \pm \eta^{2} \right)/2.
     \label{eq:3no14}
\end{eqnarray}
Here it is assumed that all neutrinos can be emitted in the muon 
decay. Then, the unitarity relations for the lepton 
mixing matrices $U$ and $V$, namely, ${\mathrm{\Sigma}}_{j}|U_{\ell j}|^2 
= {\mathrm{\Sigma}}_{j}|V_{\ell j}|^2=1$ have been used.
\par
On the other hand, in the Majorana neutrino case, the coefficients are:
\begin{eqnarray}
&&a_{\pm} 
     =\left[ \left( 1 - \overline{u_{e}}^{\, 2} \right)
     \left( 1 - \overline{u_{\mu}}^{\, 2} \right)
     \pm \lambda^2 \, \overline{v_{e}}^{\, 2} \,
     \overline{v_{\mu}}^{\, 2} \right] ,
     \label{eq:3no16} \nonumber \\
&&k_{\pm \, c}   
     =\left[ \kappa^2 \, (1 - \overline{u_{e}}^{\, 2} )
      \, \overline{v_{\mu}}^{\, 2}
      \pm \eta^{2} \, \overline{v_{e}}^{\, 2} \,
      (1 - \overline{u_{\mu}}^{\, 2} ) \right]/2 ,
     \label{eq:3no17} \\
&&k_{\pm \, m} 
     =\left[ \,\kappa^2 \,
     | \, \overline{w_{e \mu}} \, |^{2}
     \pm \eta^{2} \,
     | \, \overline{w_{e \mu \, h}} \, |^{2} \, \right]/2 \, ,
     \label{eq:3no19} \nonumber
\end{eqnarray}
where $\overline{u_{\ell}}^{\, 2}$, $\overline{v_{\ell}}^{\, 2}$, 
$\overline{w_{e  \, \mu}}$ and $\overline{w_{e \, \mu \, h}}$ 
are small quantities. 
Note that $k_{\pm \, c}$ has the same order of 
magnitude as $k_{\pm \, m}$, in contrast to the Dirac neutrino 
case. In the Majorana neutrino case, 
we assume the existence of heavy Majorana neutrinos, 
which are not emitted in the muon decay. 
Then, we have ${\mathrm{\Sigma}}_{j}^{ \, \prime}|U_{\ell j}|^2 
 = 1-\overline{u_{\ell}}^{\, 2}$ and 
${\mathrm{\Sigma}}_{j}^{ \, \prime}|V_{\ell j}|^2 
 = \overline{v_{\ell}}^{\, 2}$
where the primed sums are taken over only the light 
neutrinos. In addition, the following products 
of $U$ and $V$ appear:
$\overline{w_{e \mu}}
  \equiv {\mathrm{\Sigma}}_{j}^{\, \prime} \, U_{ej} \, V_{\mu j}
     \quad \mbox{and} \quad
\overline{w_{e \mu \, h}}
  \equiv {\mathrm{\Sigma}}_{k}^{\, \prime} \, V_{ek} \, U_{\mu k}$.
%%%%%%%%%%%%%%%%%%%%%%%%%%%%%%%%%%%%%%%%%%%%%%%%%%%%%%%%%%%%%%%%%%%%%%%%%%%%
\section{Energy spectrum of $e^{+}$ in the $\mu^{+}$ decay}
We propose a new parameterization that directly represents deviations 
from the standard model. Namely, 
if we assume the $SU(2)_{L} \times SU(2)_{R} \times U(1)$ model 
for simplicity, our expression for the energy spectrum of $e^{+}$ becoms 
\begin{eqnarray}
xD(x, \theta) 
  = x^2 [ (3 - 2 x) &+& 2 \rho_{c} (3 - 4 x)
     +  12 \varepsilon_{m} \rho_{m}  (1 - x)]  \nonumber \\
     &+& x^2 P_{\mu} \xi \cos \theta ( - 1 + 2 x ), \label{eq:3no42}
\end{eqnarray}
where the parameters $\rho_{c}$, $\rho_{m}$, and $\xi$ are, respectively, 
given by
\begin{equation}
\rho_{c} = \frac{k_{+ \, c}}{A_{1 \, 0}}
     > 0, \hspace{3mm} \rho_{m} =
      \frac{k_{+ \, m}}{A_{1 \, 0}}  > 0,
     \hspace{3mm} \mbox{and} \hspace{3mm}
     \xi =
      \frac{a_{-}+ 6 k_{- \, c}}{A_{1 \, 0}}, 
\end{equation}
with the choice of normalization factor 
$A_{1 \, 0} = a_{+} + 2 \, k_{+ \, c}$. 
Note that we have
\begin{eqnarray}
\xi &=& \frac {1 - \lambda^{2}}
     {1 + \lambda^{2} + 2 \, \eta^{2}}
     \hspace{14mm} \mbox{for the Dirac neutrino case,} 
     \label{eq:3no37} \\
\xi &\simeq& 1
     \hspace{35mm} \mbox{for the Majorana neutrino case,}
     \label{eq:3no38} \\
\xi &=& 1
     \hspace{35mm} \mbox{for the standard model.}
\end{eqnarray}
%%%%%%%%%%%%%%%%%%%%%%%%%%%%%%%%%%%%%%%%%%%%%%%%%%%%%%%%%%%%%%%%%%%%%%%%%%%%%
\subsection{Dirac type neutrino case}
Since the well-known Michel parameter $\rho_{M}$ is related to our 
$\rho_{c}$ as $2 \rho_{c} = \left( 1- \frac{4}{3}\rho_{M} \right)$, 
we have the following expression:
\begin{eqnarray}
xD(x, \theta)
     &=& x^2 \left[  (3 - 2 x) + 2 \rho_{c} (3 - 4 x) 
       + P_{\mu} \, \xi \, \cos \theta \,( - 1 + 2 x ) \right] \\
     &=& 6 x^2 [ (1 - x) + \frac{2}{9} \rho_{M} (4 x-3)] 
       + P_{\mu} \, \xi \, \cos \theta \,x^2( - 1 + 2 x ).
\end{eqnarray}

The TWIST group~\cite{TwistRho} reported a precise experimental result 
\begin{eqnarray}
  \rho_{M} = 0.75080 \pm 0.00032\mathrm{(stat)} \pm
   0.00097\mathrm{(syst)} \pm 0.00023,
     \label{eq:3no20}
\end{eqnarray}
by using angle-integrated spectrum derived from Eq.(6.7). 
The mean value is $\rho_{M}=0.75080$, although $\rho_{M} < 0.75$ is 
satisfied within experimental uncertainty.
Note that theoretical prediction from our model is 
$\rho_{M}=0.75(1-2\rho_{c})<0.75$, because
$\rho_{c} = \eta^2 / (1 + \lambda^2 + 2 \eta^2)>0$. 
They determined this $\rho_{M}$ by minimizing the $\chi^{2}$ value.  
%%%%%%%%%%%%%%%%%%%%%%%%%%%%%%%%%%%%%%%%%%%%%%%%%%%%%%%%%%%%%%%%%%%%%%%%%%%%%
\subsection{Majorana type neutrino case}
The following expression is derived for the Majorana type neutrino case:
\begin{equation}
xD(x, \theta) 
  = x^2 \left[ (3 - 2 x) + 2 \rho_{c} (3 - 4 x) +12 \rho_{m} (1-x)
  +P_{\mu} \xi \cos \theta ( - 1 + 2 x )\right].
\end{equation}
Theoretically we predict
\begin{eqnarray} 
\rho_{c} \simeq \frac{1}{2}
   \left( \kappa^2 \, \overline{v_{\mu}}^{\, 2}
   + \eta^{2} \, \overline{v_{e}}^{\, 2} \right) > 0,\hspace{2mm} 
\rho_{m} \simeq \frac{1}{2}
   \left( \kappa^2 \, |\overline{w_{e \mu}}|^2
   + \eta^{2} \, |\overline{w_{e \mu \, h}}|^2 \right) > 0.
\end{eqnarray}
\par
Therefore, there is a possibility to take advantage of the 
different $x$ dependences of the terms including the 
parameters $\rho_{c}$ and $\rho_{m}$ in the energy spectrum, when 
$\chi^{2}$ value for the Dirac-type neutrino case is compared 
with the one for the Majorana-type neutrino case. 
This may provide a test 
to determine the type of neutrino, although it is indirect.

%%%%%%%%%%%%%%%%%%%%%%%%%%%%%%%%%%%%%%%%%%%%%%%%%%%%%%%%%%%%%%%%%%%%%%%%%%%%
\section{Summary: How to determine the type of neutrino}
\par 
There is a method that might make it possible to 
distinguish between two neutrino types.  
This method makes use of the different $x$ dependences of the terms 
including the parameters $\rho_{c}$ and $\rho_{m}$ in the energy spectrum 
by comparing the $\chi^{2}$ values for the Dirac-type neutrino with those 
for the Majorana-type neutrino. Here we use the data with 
$\theta=\pi/2$, so that it may need new data-taking. 
\par
For example, suppose we analyze experimental data with $\theta=\pi/2$ 
by using 
\begin{equation}
xD(x, \theta=\pi/2) 
  = x^2 \left[ (3 - 2 x) + 2 \rho_{c} (3 - 4 x) +12 \rho_{m} (1-x)\right],
\end{equation}
and obtain some $\chi^{2}$ value, say, $\chi^{2}_{m}$ for 
the Majorana neutrino case. 
Then, suppose we repeat a similar analysis using 
\begin{equation}
xD(x, \theta=\pi/2) 
  = x^2 \left[ (3 - 2 x) + 2 \rho_{c} (3 - 4 x) \right],
\end{equation}
and thereby determine $\chi^{2}_{d}$ for the Dirac neutrino case.  
\par
If $\chi^{2}_{m}$ is  much smaller than $\chi^{2}_{d}$, 
we can conclude that there is a higher probability that neutrinos are of 
the Majorana type.

\begin{acknowledgments}
The author would like to thank Prof.~Tsuneyuki Kotani and Prof.~Masaru Doi 
for very helpful discussions in preparing this manuscript.
\end{acknowledgments}

\appendix%*
%%%%%%%%%%%%%%%%%%%%%%%%%%%%%%%%%%%%%%%%%%%%%%%%%%%%%%%%%%%%%%%%%%%%%%%%%%%%
\section{normalization factor $A$ and the $\chi^{2}$-fitting}
We summarize the meaning of the freedom for the normalization 
factor $A$ in the $\chi^{2}$-fitting of the $e^{+}$ energy 
spectrum $x D(x)$. 
\par
First let us consider the case where $x D(x)$ consists of  
the leading term $f(x)$ and its small deviation term $g(x)$ with weight 
$w$, namely,
\begin{equation}
x D(x)=\frac{1}{A} \left[f(x)+w \, g(x)\right].
\end{equation}
Then the unknown constant $w$ is determined experimentally by minimizing 
the $\chi^{2}_{\mathrm{A}}$-value~\cite{Cowan}, that is,
\begin{eqnarray}
\chi^{2}_{\mathrm{A}} =
     {\mathrm{\Sigma}}_{k} \left[
     \frac{d(x_{k}) - y_{\mathrm{A}, \, k}}{\sigma_{k}} \right]^{2},
\end{eqnarray}
where $x_{k}$ stands for the energy of the $k$-th positron 
observed by experiment, 
and $d(x_{k})$ and $\sigma_{k}$ mean, respectively, the corresponding 
experimental value of spectrum and its experimental error. 
The theoretical expression is 
$y_{\mathrm{A}, \, k} = f(x_{k})+w \, g(x_{k})$ in this case. 
The summation is taken over all observed data. 
This case is referred to as analysis A, which corresponds to the 
choice of the normalization factor $A=1$. 
\par
Since we have 
$\chi^{2}_{\mathrm{A}}
     = a_{\mathrm{A}} w^{2} - 2 b_{\mathrm{A}} w + c_{\mathrm{A}}$ in 
this analysis A,
the minimal value of $\chi^{2}_{\mathrm{A}}$ and the corresponding 
value of $w$, respectively, are determined by 
\begin{eqnarray}
\chi^{2}_{\mathrm{A}, \, min} =
     \left( c_{\mathrm{A}} -
             \frac{b_{\mathrm{A}}^{2}}{a_{\mathrm{A}}} \right)
   \quad \mbox{and} \quad
     w_{\mathrm{A}} = \left( \frac{b_{\mathrm{A}}}{a_{\mathrm{A}}}
     \right).
\end{eqnarray}
Here $a_{\mathrm{A}}$, $b_{\mathrm{A}}$, and $c_{\mathrm{A}}$ are defined by 
\begin{eqnarray}
a_{\mathrm{A}} &=& {\mathrm{\Sigma}}_{k}
     \left( \frac{g(x_{k})}{\sigma_{k}} \right)^{2},
     \quad \quad
  b_{\mathrm{A}} = {\mathrm{\Sigma}}_{k}
     \left( \frac{g(x_{k})}{\sigma_{k}} \right)
     \left( \frac{d(x_{k}) - f(x_{k})}{\sigma_{k}} \right),
     \nonumber \\
  c_{\mathrm{A}} &=& {\mathrm{\Sigma}}_{k}
     \left( \frac{d(x_{k}) - f(x_{k})}{\sigma_{k}} \right)^{2}.
\end{eqnarray}
\par
Next, let us consider the case where terms in $x D(x)$ are rearranged
as follows,
\begin{equation}
 x D(x)=\frac{1}{A} \left[f(x)+w \, g(x)\right]=\left(\frac{1+w}{A} \right) \left\{ f(x)+\frac{w}{1+w}
     [g(x)-f(x)] \right\}. 
\end{equation}
Then we can determine $\left( \frac{w}{1+w} \right)$ 
experimentally by minimizing the $\chi^{2}_{\mathrm{B}}$-value,
\begin{eqnarray}
\chi^{2}_{\mathrm{B}} =
     {\mathrm{\Sigma}}_{k} \left[
     \frac{d(x_{k}) - y_{\mathrm{B}, \, k}}{\sigma_{k}} \right]^{2},
\end{eqnarray}
where the theoretical expression is defined by
\begin{equation}
y_{\mathrm{B}, \, k} = f(x_{k})+\frac{w}{1+w} \, [g(x_{k})-f(x_{k})]. 
\end{equation}
This case is referred to as analysis B, which corresponds to the choice 
of the normalization factor $A=(1+w)$. 
The minimal value of $\chi^{2}_{\mathrm{B}}$ and the unknown constant 
$w$ in this analysis B become as follows
\begin{eqnarray}
\chi^{2}_{\mathrm{B}, \, min} =
     \left( c_{\mathrm{B}} -
             \frac{b_{\mathrm{B}}^{2}}{a_{\mathrm{B}}} \right),
   \quad \mbox{and} \quad
     w_{\mathrm{B}}
     = \left( \frac{b_{\mathrm{B}}}{a_{\mathrm{B}} - b_{\mathrm{B}}}
     \right).
\end{eqnarray}
Here $a_{\mathrm{B}}$, $b_{\mathrm{B}}$, and $c_{\mathrm{B}}$ are defined by  
\begin{eqnarray}
  a_{\mathrm{B}} &=& {\mathrm{\Sigma}}_{k}
     \left( \frac{g(x_{k}) - f(x_{k})}{\sigma_{k}} \right)^{2},
     \nonumber \\
  b_{\mathrm{B}} &=& {\mathrm{\Sigma}}_{k}
     \left( \frac{g(x_{k}) - f(x_{k})}{\sigma_{k}} \right)
     \left( \frac{d(x_{k}) - f(x_{k})}{\sigma_{k}} \right),
     \quad \quad
  c_{\mathrm{B}} = c_{\mathrm{A}}.
\end{eqnarray}
\par
In general, the value of $w_{\mathrm{B}}$ is different from 
$w_{\mathrm{A}}$.  Therefore, the final value of $w$ determined 
experimentally should be chosen as the $w_{\mathrm{B}}$, if 
$\chi^{2}_{\mathrm{A}, \, min} > \chi^{2}_{\mathrm{B}, \, min}$, 
or vice versa. 
So, this means that there is freedom in the $\chi^{2}$-fitting 
such as the choice of the normalization factor $A$ 
in the analysis of $xD(x)$.  
\par
It is worthwhile to note that the $x$ dependence 
of $[d(x) - f(x)]$ plays an important role in the above procedure. 
That is, in the muon decay, it means the $x$ dependence of the 
difference between the experimental data 
and the prediction from the standard model. 
Let us consider the following imaginative example:  If the 
$x$ dependence of $[d(x) - f(x)]$ happens to be proportional to that 
of $g(x)$, that is, $[d(x) - f(x)] = L_{\mathrm{A}} g(x)$, then 
we have $\chi^{2}_{\mathrm{A}, \, min} = 0$ and correspondingly 
$w_{\mathrm{A}} = L_{\mathrm{A}}$ in analysis A.  In other words, 
we have the relation, 
$\chi^{2}_{\mathrm{A}, \, min} \ll \chi^{2}_{\mathrm{B}, \, min}$. 
While, if $[d(x) - f(x)] = L_{\mathrm{B}} [g(x)-f(x)]$, then 
we have $\chi^{2}_{\mathrm{B}, \, min} = 0$ and correspondingly 
$w_{\mathrm{B}} = L_{\mathrm{B}} / (1 - L_{\mathrm{B}})$ in analysis 
B.  Namely, we have the opposite conclusion, 
$\chi^{2}_{\mathrm{A}, \, min} \gg \chi^{2}_{\mathrm{B}, \, min}$.
%%%%%%%%%%%%%%%%%%%%%%%%%%%%%%%%%%%%%%%%%%%%%%%%%%%%%%%%%%%%%%%%%%%%%%%%%%%%

%%%%%%%%%%%%%%%%%%%%%%% ref %%%%%%%%%%%%%%%%%%%%%%%%%%%%%%%%%%%%%%%%%%%%%%%

%%%%%%%%%%%%%%%%%%%%%%%%%%%%%%%%%%%%%%%%%%%%%%%%%%%%%%%%%%%%%%%%%%%%%%%%%%%%

\end{document}